\documentclass{article}

\usepackage{amsmath}

\PassOptionsToPackage{numbers, compress}{natbib}

\usepackage[preprint]{neurips_2024}
\usepackage{soul}

\usepackage[utf8]{inputenc} 
\usepackage[T1]{fontenc}    
\usepackage{hyperref}       
\usepackage{url}            
\usepackage{booktabs}       
\usepackage{amsfonts}       
\usepackage{nicefrac}       
\usepackage{microtype}      
\usepackage{xcolor}         
\usepackage{graphicx}
\usepackage{makecell}
\usepackage{multirow} 
\usepackage{natbib}

\title{Non-Cooperative Backdoor Attacks in Federated Learning: A New Threat Landscape}

\author{%
  Tuan Nguyen$^{1,2}$, Dung Thuy Nguyen$^{3}$, Khoa D Doan$^{1, 2}$, Kok-Seng Wong$^{1, 2}$\\
  $^1$VinUni-Illinois Smart Health Center, VinUniversity, Hanoi, Vietnam\\
  $^2$College of Engineering \& Computer Science, VinUniversity, Hanoi, Vietnam\\
  $^3$
  Vanderbilt University, Nashville, TN 37235\\
  \texttt{\{tuan.nm, khoa.dd, wong.ks\}@vinuni.edu.vn},\\
  \texttt{dung.t.nguyen@Vanderbilt.Edu}\\
}

\begin{document}

\maketitle

\begin{abstract}
Despite the promise of Federated Learning (FL) for privacy-preserving model training on distributed data, it remains susceptible to backdoor attacks. These attacks manipulate models by embedding triggers (specific input patterns) in the training data, forcing misclassification as predefined classes during deployment. Traditional single-trigger attacks and recent work on cooperative multiple-trigger attacks, where clients collaborate, highlight limitations in attack realism due to coordination requirements. We investigate a more alarming scenario: non-cooperative multiple-trigger attacks. Here, independent adversaries introduce distinct triggers targeting unique classes. These parallel attacks exploit FL's decentralized nature, making detection difficult. Our experiments demonstrate the alarming vulnerability of FL to such attacks, where individual backdoors can be successfully learned without impacting the main task. This research emphasizes the critical need for robust defenses against diverse backdoor attacks in the evolving FL landscape. While our focus is on empirical analysis, we believe it can guide backdoor research toward more realistic settings, highlighting the crucial role of FL in building robust defenses against diverse backdoor threats. The code is available at \url{https://anonymous.4open.science/r/nba-980F/}.

\end{abstract}

\section{Introduction}

Federated learning (FL)~\cite{mcmahan2017communication} is a distributed machine learning paradigm that enables multiple parties to train a shared model cooperatively without sharing their private data. In FL, each party trains a local model on its data and then shares the model parameters with a central server. The server aggregates the parameters from all parties and updates the global model. The updated global model is then sent back to each party for further training. This process is repeated until the global model converges.
However, because the training data is distributed across multiple parties, FL is vulnerable to backdoor attacks~\cite{nguyen2024backdoor}, where the attacker poisons the model by injecting a backdoor trigger into the training data. When the model is deployed, a specific input pattern can activate the backdoor trigger to cause the model to output a specific target class.

Backdoor attacks on FL have been recently studied in~\cite{bagdasaryan2020backdoor, wang2020attack, xie2020dba, gong2022coordinated, zhang2022neurotoxin, dai2023chameleon, nguyen2023iba}. 
Existing research categorizes these attacks into two main types: fixed-trigger attacks and optimized-trigger attacks. Fixed-trigger attacks, as described in~\cite{bagdasaryan2020backdoor}, involve pre-selecting a trigger without leveraging information from the FL training process. 
Conversely, optimized-trigger attacks refine the trigger specifically to enhance the attack's effectiveness by utilizing such information. Recent studies have explored various optimization techniques: maximizing the difference between clean and trigger-added sample representations (Fang et al., 2023)~\cite{fang2023vulnerability}, jointly optimizing the trigger and local model with regularization to bypass defenses (Lyu et al., 2023)~\cite{lyu2023poisoning}, and using autoencoders to generate the optimal trigger pattern (Nguyen et al., 2023)~\cite{nguyen2023iba}.
However, these works primarily focus on a cooperative attack scenario where malicious clients either coordinate with decomposed triggers and target labels (Xie et al., 2020)~\cite{xie2020dba} or embed the same global trigger pattern across all attackers (Bagdasaryan et al., 2020)~\cite{bagdasaryan2020backdoor}, ultimately compromising the global model.

Recent advancements in machine learning have led to new backdoor attack scenarios in FL. For instance, attackers can choose an arbitrary target class during inference (Doan et al., 2022)~\cite{doan2022marksman} or inject multiple triggers to poison the same dataset (Li et al., 2024)~\cite{li2024multi}. These advancements highlight the attacker's growing sophistication, allowing independent attackers to inject their triggers and target classes without coordination. This approach is more realistic and poses a significant challenge for real-world FL systems, as any participant can potentially learn a backdoor task without compromising the main task's performance.
Independent attacks can be motivated by individual goals. Imagine competing companies participating in an FL system to develop a recommendation model. A malicious company could attempt to inject a backdoor into the model to promote its own products to users unfairly.
Motivated by this emerging threat, we investigate a new backdoor attack scenario in FL: Non-Cooperative Backdoor Attacks (NBA). In this scenario, adversarial clients act independently, each with a unique backdoor trigger and target class. This scenario presents a significant practical challenge, as it allows any participant to introduce a backdoor without affecting the core functionality of the model. Fig.~\ref{fig:non-cooperative-attack} provides a visual overview of the proposed NBA.

Through extensive experiments, we summarize our main contributions as follows:
\begin{itemize}
    \item We introduce a new attack scenario: Non-Cooperative Backdoor Attacks (NBA) in FL, where multiple malicious clients act independently by employing their specific trigger to backdoor their own targeted class. This scenario reflects a more realistic threat in real-world FL deployments.
    
    \item We demonstrate the efficacy and increased risk of NBA attacks through extensive experiments on four datasets. We show successful backdoor insertion in single-shot, multiple-shot, and semi-multiple-shot settings, highlighting the growing danger as attackers exploit multiple communication rounds.
    
    \item Our analysis investigates NBA with a large-scale attacker pool (up to 8 attackers). This reflects the potential for multiple parties to inject backdoors into a single model, a significant concern for practical FL systems.
    
    \item We conduct in-depth analysis and ablation studies to understand how various factors like trigger patterns, scaling factors, and the number of attackers impact NBA success. This comprehensive analysis provides valuable insights for designing future defenses.
\end{itemize}

\section{Background and related work}

\subsection{Federated learning}
In FL, users with private data collaborate to train a global model ($G^t$). Each user iteratively updates a local model ($W_i^{t+1}$ ) based on their data ($D_i$) using Stochastic Gradient Descent:
\begin{equation}{W_{i}^{t+1}} = {G^t} - lr \cdot \nabla {\mathcal{L}_{task}}({G^t},{\mathcal{D}_i})\tag{1}\end{equation}

where $\mathcal{L}_{task}$ is the loss function, $\nabla {\mathcal{L}_{task}}(G^t, D_i)$ denotes the gradient, and $lr$ is the local learning rate. User updates are then uploaded and aggregated by the central server using aggregation rules (e.g., FedAvg~\cite{mcmahan2017communication}) with a global learning rate ($\eta$) to create a new global model for the next round:
\begin{equation}
{G^{t + 1}} = {G^t} + \frac{\eta }{n}\mathop \sum \limits_{i = 1}^n ({W_{i}^{t+1} } - {G^t})\tag{2}
\end{equation}

\subsection{Backdoor attacks and defenses in FL}

\textbf{Backdoor attack. } This attack aims to make a model perform well on normal data (benign data) while also producing attacker-desired outputs for inputs with a hidden trigger (e.g., specific image pattern). Attackers participate in FL with backdoor data ($D_{backdoor}$ ) to poison the model. Eq. \ref{eq:backdoor_model_update} shows how the attacker updates their local model ($W_{adv}^{t+1}$ ) using both normal data ($D_{normal}$ ) and backdoor data:
\begin{equation}
\label{eq:backdoor_model_update}
W_{adv}^{t+1} = G^t - lr \cdot \nabla \mathcal{L}_{task}(G^t, D_{normal} \cup D_{backdoor}) \tag{3}
\end{equation}

\textbf{Existing backdoor attacks in FL.} A widely recognized fact is that a global model approaching convergence undergoes minimal significant gradient updates. In light of this, attackers can employ a scaling factor $\gamma$ in the Model Replacement Attack~\cite{bagdasaryan2020backdoor} to replace the global model with an attacker-trained backdoor model within a single epoch. Leveraging the distributed nature of FL, the Distributed Backdoor Attack (DBA)\cite{xie2020dba} decomposes the global trigger pattern into multiple local triggers, assigning each compromised device a unique local trigger. Building upon DBA, Gong et al.\cite{gong2022coordinated} introduce a coordinated backdoor attack with model-dependent local triggers. The semantic backdoor, a variant of the FL backdoor attack, incorporates triggers related to inherent features in target images, such as the color of a car~\cite{bagdasaryan2020backdoor}. Success and persistence in a semantic backdoor hinge on the frequency of trigger features in other clients' datasets, as highlighted by Bagdasaryan et al. To enhance backdoor effectiveness, Wang et al.\cite{wang2020attack} propose an edge-case backdoor, similar to the semantic backdoor but strategically positions the backdoor datasets at the tail of the global datasets' distribution, making them less likely to appear on other clients' data. For increased backdoor durability, Neurotoxin~\cite{zhang2022neurotoxin} identifies parameters infrequently updated by benign clients and inserts backdoors using these parameters. Chameleon~\cite{dai2023chameleon} explores the relationship between the original label and the backdoor label before flipping the label to the target label to extend the backdoor duration. IBA~\cite{nguyen2023iba} leverages the updated history of adversarials with imperceptible triggers to enhance backdoor durability.

\textbf{Existing backdoor defenses in FL.} Existing defense methods primarily focus on distinguishing adversarial' updates from benign clients' updates, as adversarial strive to make their updates closely resemble other updates. 
Various outlier detection techniques have been proposed to counter backdoor attacks. Li et al. proposed a spectral anomaly detection framework based on low-dimensional embeddings, removing noisy and irrelevant features while retaining essential ones~\cite{li2020learning}. In this low-dimensional latent feature space, abnormal (malicious) model updates can be easily differentiated from normal updates.
Deepsight~\cite{rieger2022deepsight} conducts deep model inspection for each model, analyzing Normalized Update Energies (NEUPs) and Division Differences (DDifs). FL-Detector~\cite{zhang2022fldetector} predicts the global model through model update consistency, detecting outliers based on the distance to the predicted model. RFLBAT~\cite{wang2022rflbat} utilizes Principal Component Analysis (PCA) to reduce the dimension of gradient updates, effectively separating malicious models from benign models in a low-dimensional projection space.
Foolsgold~\cite{fung2020limitations}, examines historical updates for each client and penalizes those with high pairwise cosine similarities by employing a low learning rate. 
Another avenue of research focuses on robust defense against FL backdoor attacks by applying weak Differential Privacy (DP)~\cite{zheng2020protecting} to the global model~\cite{zheng2020preserving}. Weak DP, involving norm clipping and the addition of Gaussian noise to each gradient update, has proven effective in mitigating FL backdoor attacks~\cite{xie2021crfl}. Recognizing potential drawbacks such as deteriorating the global model's main task accuracy with Gaussian noise and the need for a clipping bound with norm clipping, FLAME adapts DP method by introducing a noise boundary proof and a dynamic clipping bound. This adaptation has demonstrated its capability to alleviate backdoor attacks while still maintaining a high main task accuracy~\cite{nguyen2022flame}.

\noindent \textbf{Knowledge of the adversary.}  While the attacker has white-box access to the global model weights and predictions, their knowledge of the training data is limited to the data distribution held by compromised clients, resulting in only partial knowledge of the overall training data.

\textbf{Capabilities of the adversary.}  We assume a model-poisoning adversary with full access to the server and a fixed number of compromised clients, similar to the scenario presented in Xie et al.~\cite{xie2020dba}. This powerful attacker can alter training hyperparameters, model weights, and training data on compromised clients.

\section{Non-Cooperative Backdoor Attacks against federated learning}
\label{sec:framework}
\subsection{General framework}

\begin{figure}[!hbt]
    \centering
    \includegraphics[width=1\linewidth]{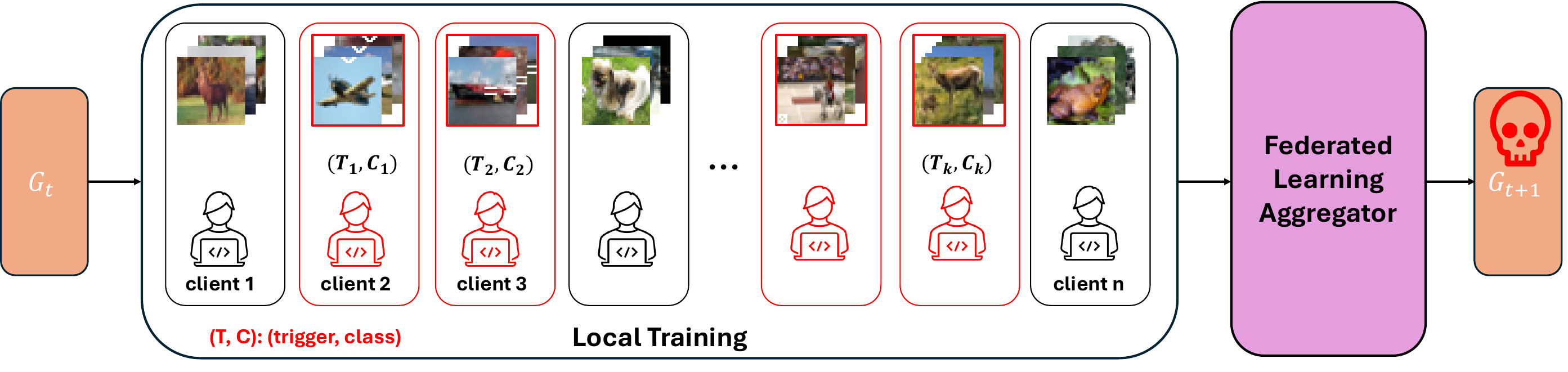}
    \caption{Non-Cooperative Backdoor Attacks (NBA) scenario in FL: the red color represents the malicious client with their own unique trigger and target class, aiming to inject the backdoor trigger. Here, $(T_i, C_i)$ denotes the trigger and target class of the $i$-th attacker.}
    \label{fig:non-cooperative-attack}
\end{figure}

Our proposed scenario, NBA, in FL, involves multiple clients acting independently, each with their own unique backdoor trigger and target class.
As shown in Fig.~\ref{fig:non-cooperative-attack}, this attack differs from from existing cooperative backdoor attacks, where clients coordinate with decomposed triggers and target labels, ultimately compromising the global model. In NBA, each client acts independently, introducing its own unique backdoor trigger and target class. This novel scenario presents a significant threat to FL systems, as individual backdoor tasks can be successfully learned without harming the main task performance. The triggers and target classes of the attackers are denoted as $(T_i, C_i)$, where $i$ is the index of the attacker. All the triggers (Fig.~\ref{fig:trigger-list}) are unique in shape and have a size of 24 pixels~\cite{xie2020dba}.

\subsection{Factors in Non-Cooperative Backdoor Attacks}

\begin{figure}[!hbt]
    \centering
    \includegraphics[width=1.0\linewidth]{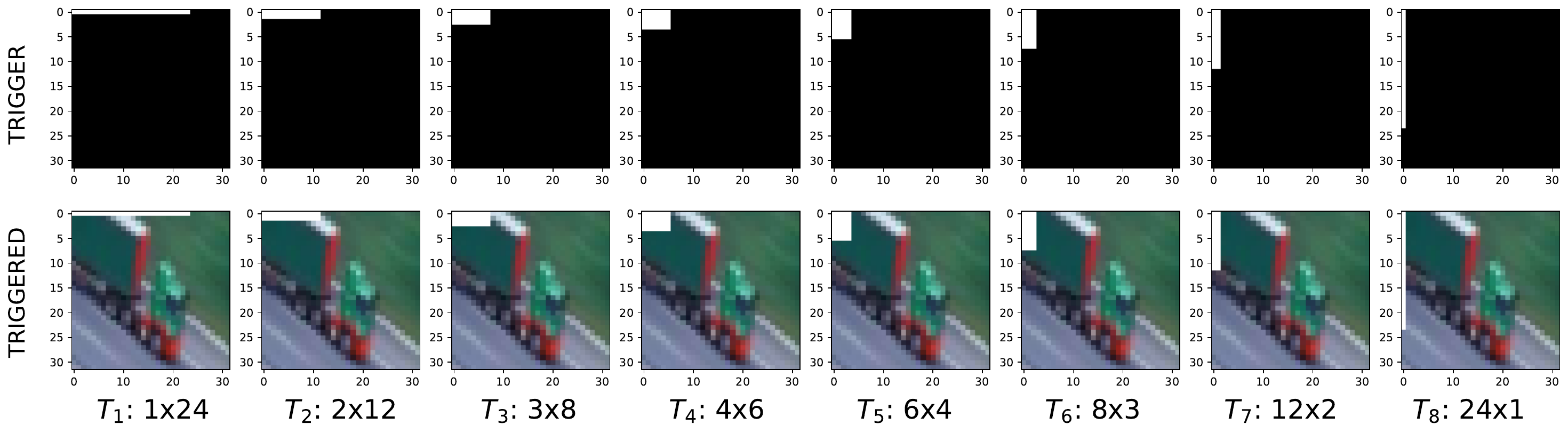}
    \caption{Eight trigger patterns used in our NBA experiments, all with fixed sizes of 24 pixels.}
    \label{fig:trigger-list}
\end{figure}

\textbf{Trigger location and size.} The trigger is located at the top left corner of the image, with a size of 24 pixels. We use eight fixed trigger patterns with the shapes $1 \times 24$, $2 \times 12$, $3 \times 8$, $4 \times 6$, $6 \times 4$, $8 \times 3$, $12 \times 2$, and $24 \times 1$ pixels, respectively, as shown in Fig.~\ref{fig:trigger-list}.

\textbf{Scale $\gamma$.} The scaling parameter $\gamma = \frac{n}{\eta}$ defined in Bagdasaryan et al.~\cite{bagdasaryan2020backdoor} is used by the attacker to scale up the malicious model weights. For instance, assume the ith malicious local model is $X$. The new local model $L_{i}^{t+1}$ that will be submitted is calculated as $L_{i}^{t+1} = \gamma (X - G_t) + G_t$.

\textbf{Data distribution $\alpha$.} FL often presumes non-i.i.d. data distribution across parties. Here, we use a Dirichlet distribution~\cite{Minka00estimatinga} with hyperparameter $\alpha = 0.5$ to generate different data distributions.

\section{Experiments}
\label{sec:experiments}

\subsection{Datasets and experiment setup}
\label{subsec:dataset-exp-setup}
\textbf{Datasets.}  NBA is evaluated on four classification datasets with non-i.i.d. data distributions: Fashion-MNIST~\cite{xiao2017fashion}, MNIST~\cite{lecun1998gradient}, CIFAR-10~\cite{krizhevsky2009learning}, and Tiny-ImageNet~\cite{le2015tiny}.
The data description and parameter setups are summarized in Appx.~\ref{appsub:exp_setup}.

\textbf{Federated learning setup.}
Following the standard setup, we use FedAvg~\cite{mcmahan2017communication} as the global model optimization algorithm, and the global learning rate $\eta$ is set to 0.01. In each round, 10 of the 100 clients are selected for aggregation and each selected client trains for $E$ local epochs with a local learning rate $lr$. Our experiments utilize 8 triggers with fixed sizes of 24 pixels, as shown in Fig.~\ref{fig:trigger-list}.

\textbf{Attack scenarios.} We evaluate the performance of NBA in three distinct attack scenarios following the setup in~\cite{bagdasaryan2020backdoor, xie2020dba}:
\begin{itemize}
    \item \textit{Single-shot attack:} Attackers participate in only one round, during which they scale the client's model using the model replacement method~\cite{bagdasaryan2020backdoor} with a scaling factor of $\gamma = 100$.
    \item \textit{Multiple-shot attack:} Attackers are continuously selected to participate throughout the entire training process, without applying any scaling factor.
    
    \item \textit{Semi-multiple-shot attack:} Attackers are continuously selected for a fixed number of rounds (100 rounds in our experiments). They employ the model replacement method combined with varying scaling factors $\gamma$ ranging from 1 to 100, blending the continuous participation aspect of~\cite{xie2020dba} with the scaling strategy of~\cite{bagdasaryan2020backdoor}.
\end{itemize}

It is important to note that all attack settings are initiated after the global model has converged. Injecting backdoors from the first round, as observed in Xie et al.~\cite{xie2020dba}, can lead to low main accuracy and difficulty in model convergence.

\noindent \textbf{Evaluation metrics.}
We use the following evaluation metrics to measure the performance of the proposed NBA in FL:
\begin{itemize}
    \item \textbf{Main Task Accuracy ($MA$):} Accuracy of the global model on the main task during testing.

    \item \textbf{Backdoor Task Accuracy ($BA$):} Percentage of test inputs with a specific pixel pattern correctly classified into the target class by global model. For trigger $k$, this is $BA_k$.
    
\end{itemize}

\subsection{Backdoor attacks with one adversary}

\subsubsection{Single-shot attack}

We begin our evaluation by analyzing the performance of an attack in a single-shot setting with one adversary. In this setting, the attacker participates in only one round of training and modifies the strength of the backdoor model with a scaling factor $\gamma = 100$.

\begin{figure}[!hbt]
    \centering
    \includegraphics[width=1.0\linewidth]{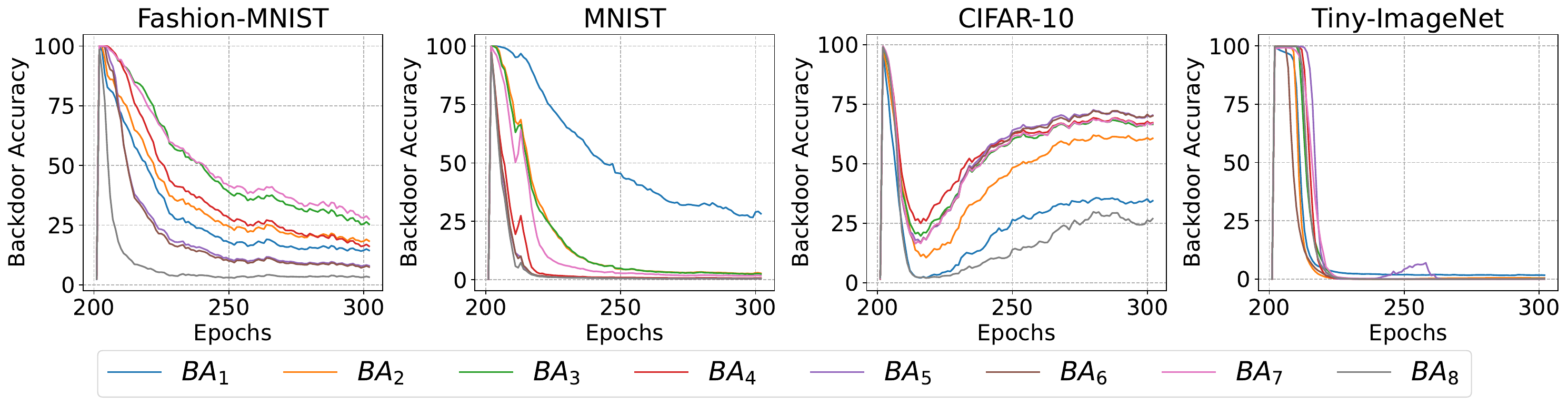}
    \caption{Backdoor accuracy of 8 triggers in single-shot attack with one adversary and $\gamma=100$.}
    \label{fig:nba-ssa-ba-100}
\end{figure}

\textbf{Performance of backdoor attacks with different triggers.} The performance of backdoor attacks with different triggers shows consistency between datasets. With $\gamma=1$, all datasets exhibit relatively low backdoor accuracies, with Fashion-MNIST and CIFAR-10 having slightly higher values compared to MNIST and Tiny-ImageNet. However, when $\gamma=100$, the backdoor accuracy is consistently high across all triggers within each dataset, indicating that the specific trigger used has less impact on the success rate when the scaling factor is high. For instance, all four datasets maintain almost perfect backdoor accuracies, close to 100\% for all triggers. This consistency suggests that the choice of trigger becomes less significant when the scaling factor is increased.

\begin{table}[!hbt]
    \centering
    \caption{Performance of backdoor attacks with one adversary in single-shot setting}
    \label{tab:backdoor-success-rate}
    \small
    \begin{tabular}{@{}lcccccccccc@{}}
    \toprule
 
\makecell{ $\gamma$} & \makecell{\textbf{Accuracy} $\rightarrow$,\\ \textbf{Dataset} $\downarrow$} & \textbf{$BA_1$} & \textbf{$BA_2$} & \textbf{$BA_3$} & \textbf{$BA_4$} & \textbf{$BA_5$} & \textbf{$BA_6$} & \textbf{$BA_7$} & \textbf{$BA_8$} & \textbf{$BA_{Avg}$} \\ \midrule

\multirow{4}{*}{1} 
   & Fashion-MNIST & 2.60 & 2.83 & 2.94 & 2.78 & 2.59 & 2.68 & 2.59 & 2.48 & 2.69 \\
    & MNIST & 0.33 & 0.34 & 0.33 & 0.33 & 0.37 & 0.37 & 0.38 & 0.37 & 0.35 \\
    & CIFAR-10 & 1.71 & 1.87 & 1.52 & 1.68 & 1.64 & 1.47 & 1.67 & 1.90 & 1.68 \\
    & Tiny-ImageNet & 0.09 & 0.08 & 0.09 & 0.07 & 0.06 & 0.05 & 0.07 & 0.08 & 0.07 \\
    \midrule
\multirow{4}{*}{100} 
    & Fashion-MNIST & 99.88 & 100 & 100 & 100 & 100 & 100 & 100 & 98.67 & 99.82 \\
    & MNIST & 100 & 99.99 & 100 & 98.61 & 97.92 & 97.73 & 99.96 & 97.28 & 98.94 \\
    & CIFAR-10 & 99.88 & 100 & 100 & 100 & 100 & 100 & 100 & 98.67 & 99.82 \\
    & Tiny-ImageNet & 99.37 & 99.67 & 99.92 & 99.83 & 99.77 & 99.89 & 99.30 & 99.83 & 99.70 \\
    \midrule

\end{tabular}
\end{table}

\textbf{Impact of scaling factor $\gamma$ on backdoor task.} The scaling factor $\gamma$ significant affects the performance of backdoor attacks. When $\gamma=1$, indicating no scaling, the backdoor accuracy across all datasets is relatively low. For instance, the backdoor accuracy for Fashion-MNIST ranges from 2.48\% to 2.94\%, with an average of 2.69\%, while for MNIST, it ranges from 0.33\% to 0.38\%, with an average of 0.35\%. CIFAR-10 shows backdoor accuracies from 1.47\% to 1.90\%, averaging 1.68\%, and Tiny-ImageNet displays very low values between 0.05\% and 0.09\%, with an average of 0.07\%. In contrast, when $\gamma=100$, the backdoor accuracy dramatically increases, reaching nearly 100\% across most triggers and datasets. For example, Fashion-MNIST and CIFAR-10 both achieve an average backdoor accuracy of 99.82\%, while MNIST and Tiny-ImageNet also show high averages of 98.94\% and 99.70\%, respectively. This stark contrast highlights the critical role of the scaling factor in determining the success of backdoor attacks in single-shot settings.

\textbf{Abnormality of backdoor performance on CIFAR-10 dataset.} 
The graph in Fig.~\ref{fig:nba-ssa-ba-100} depicts the backdoor accuracy trends over 100 rounds for the CIFAR-10 dataset compared to three other datasets. During attack rounds with \(\gamma = 100\), the backdoor accuracy initially reaches nearly 100\% across all four datasets. Following these rounds, in the absence of backdoor training injections, the backdoor accuracy generally decreases after 20 rounds. However, in the CIFAR-10 dataset, an unusual pattern emerges: the backdoor accuracy begins to increase after 20 rounds, eventually reaching nearly 70\% even after 100 rounds. In contrast, the backdoor accuracy in the other datasets continues to decline, approaching 0\% after 100 rounds. This observation suggests that the intrinsic characteristics of the CIFAR-10 dataset significantly influence the persistence of backdoor performance.

\subsubsection{Multiple-shot attack}

\begin{figure}[!hbt]
    \centering
    \includegraphics[width=1.0\linewidth]{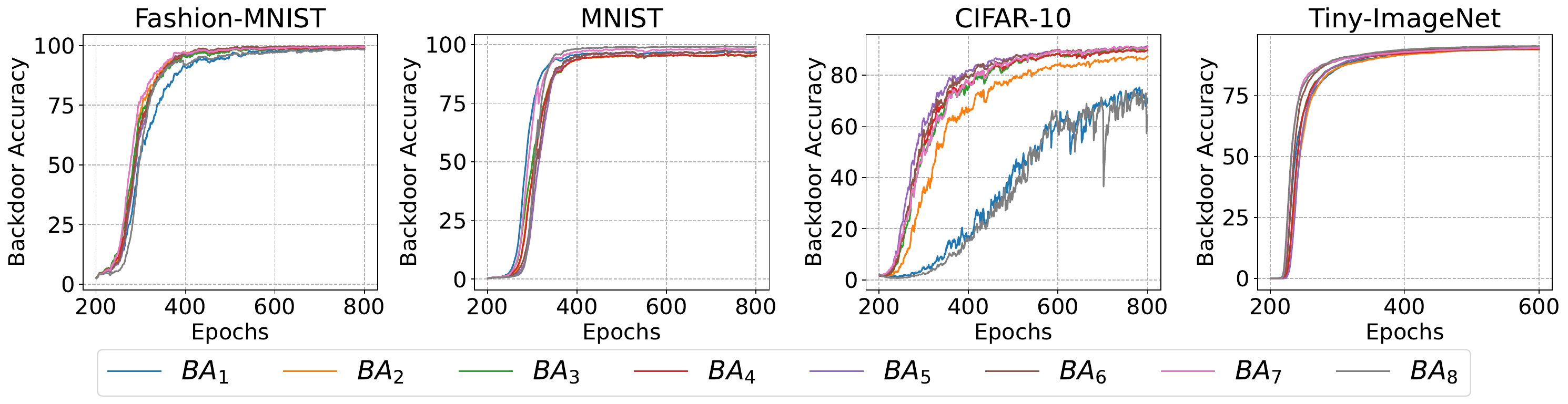}
    \caption{Backdoor accuracy in multiple-shot setting with one adversary ($\gamma=1$)}
    \label{fig:single-ba-200r}
\end{figure}

\textbf{Performance of backdoor task with different triggers.} In the multiple-shot attack scenario with one adversary over 200 rounds, as shown in Fig.~\ref{fig:single-ba-200r}, the performance of backdoor attacks exhibits notable differences across datasets. For Fashion-MNIST, the backdoor accuracy remains consistently high, with an average backdoor accuracy ($BA_{Avg}$) of 99.38\%, indicating that the backdoor attack is highly effective. Similarly, MNIST shows a strong resilience to the backdoor attack, with an average backdoor accuracy of 97.18\%, though slightly lower than Fashion-MNIST. In the case of CIFAR-10, there is a significant range in backdoor accuracy, with values spanning from 74.06\% to 91.04\%, and an average of 85.86\%, suggesting variable success in the backdoor attack across different triggers. Tiny-ImageNet displays consistently high backdoor accuracy, with all values close to or exceeding 93.80\%, culminating in an average of 94.29\%, indicating effective backdoor insertion. These results demonstrate the varying effectiveness of backdoor attacks in a multiple-shot scenario, heavily influenced by the specific triggers used.

\subsection{Non-Cooperative Backdoor Attacks}

\subsubsection{Single-shot attack}
\begin{figure}[!hbt]
    \centering
    \includegraphics[width=1.0\linewidth]{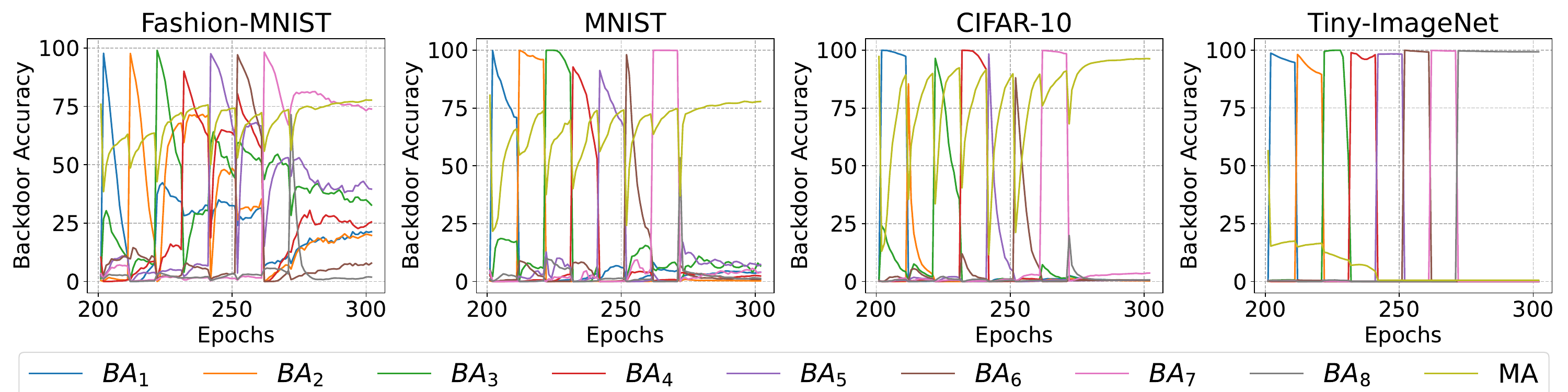}
    \caption{Backdoor accuracy in single-shot NBA with $\gamma=100$ and gap 10 rounds. }
    \label{fig:nba-single-gap-10}
\end{figure}

\begin{table}[!hbt]
    \centering
    \caption{Performance of NBA in single-shot setting with $\gamma=100$ (gap 10 rounds)}
    \label{tab:nba-single-gap-shot}
    \small
    \begin{tabular}{@{}lcccccccccc@{}}
    \toprule
 
\makecell{\textbf{Accuracy} $\rightarrow$,\\ \textbf{Dataset} $\downarrow$} & \textbf{$MA$} & \textbf{$BA_1$} & \textbf{$BA_2$} & \textbf{$BA_3$} & \textbf{$BA_4$} & \textbf{$BA_5$} & \textbf{$BA_6$} & \textbf{$BA_7$} & \textbf{$BA_8$} & \textbf{$BA_{Avg}$} \\ \midrule

Fashion-MNIST & 77.72 & 21.31 & 19.74 & 32.73 & 25.50 & 39.57 & 7.83 & 73.89 & 1.87 & 27.81 \\
MNIST & 77.84 & 3.99 & 0.28 & 6.74 & 2.36 & 7.30 & 0.97 & 3.96 & 1.29 & 3.36 \\
CIFAR-10 & 96.32 & 0.51 & 0.28 & 0.50 & 0.57 & 0.54 & 0.42 & 3.64 & 0.66 & 0.89 \\
Tiny-ImageNet & 0.50 & 0.00 & 0.00 & 0.00 & 0.00 & 0.00 & 0.00 & 0.00 & 99.30 & 12.41 \\

\bottomrule

\end{tabular}
\end{table}

Fig.~\ref{fig:nba-single-gap-10} illustrates the backdoor accuracy trends in a single-shot NBA setting with an attack gap of 10 rounds, meaning the first attacker injects the backdoor in the first round, the next in the $11^{th}$ round, and so on. In this scenario, each attacker participates in only one round, scaling the client’s model using the model replacement method with a scaling factor \(\gamma = 100\). As shown in Fig.~\ref{fig:nba-single-gap-10}, the backdoor accuracy at the scaling round is nearly 100\% for all triggers across all datasets. After the scaling round, the backdoor accuracy decreases significantly, reducing to nearly zero by the next scaling round. 
After 30 rounds without scaling, the backdoor accuracy patterns vary between datasets.  For MNIST and CIFAR-10, while the main task accuracy remains high, the backdoor accuracy for all triggers drops to nearly zero.  In the case of Fashion-MNIST, although the main task accuracy is moderate, the backdoor accuracy is varied, with some triggers maintaining higher accuracy than others.
For Tiny-ImageNet, the main task accuracy is notably affected by the backdoor attack, showing a low main task accuracy alongside a high backdoor accuracy for the last trigger.
These results highlight that dataset characteristics and backdoor trigger design are crucial factors influencing the success of backdoor attacks in FL.

\subsubsection{Multiple-shot attack}

\begin{table}[!hbt]
    \centering
    \caption{Performance of NBA in multiple-shot setting with $\gamma = 1$ (8 adversaries)}
    \label{tab:nba-multiple-shot}
    \small
    \begin{tabular}{@{}lcccccccccc@{}}
    \toprule
 
\makecell{\textbf{Accuracy} $\rightarrow$,\\ \textbf{Dataset} $\downarrow$} & \textbf{$MA$} & \textbf{$BA_1$} & \textbf{$BA_2$} & \textbf{$BA_3$} & \textbf{$BA_4$} & \textbf{$BA_5$} & \textbf{$BA_6$} & \textbf{$BA_7$} & \textbf{$BA_8$} & \textbf{$BA_{Avg}$} \\ \midrule

Fashion-MNIST & 84.07 & 86.53 & 84.11 & 86.34 & 89.68 & 95.71 & 96.90 & 97.73 & 41.26 & 84.78 \\
MNIST & 97.96 & 99.40 & 86.61 & 89.47 & 87.52 & 79.24 & 89.43 & 97.01 & 49.30 & 84.75 \\
CIFAR-10 & 76.21 & 55.87 & 88.93 & 84.69 & 86.78 & 89.99 & 86.50 & 94.09 & 27.19 & 76.75 \\
Tiny-ImageNet & 38.32 & 0.01 & 0.00 & 0.05 & 0.03 & 0.00 & 0.00 & 0.00 & 0.04 & 0.02 \\

\bottomrule
\end{tabular}
\end{table}

\begin{figure}[!hbt]
    \centering
    \includegraphics[width=1.0\linewidth]{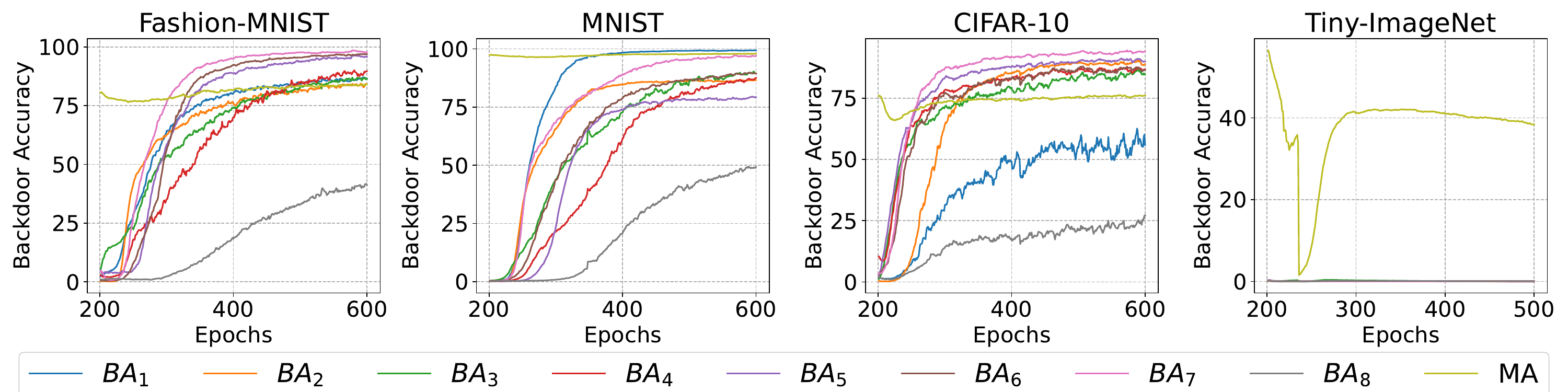}
    \caption{Performance of NBA in multiple-shot setting with 8 adversaries ($\gamma = 1$)}
    \label{fig:nba-mul-1}
\end{figure}

\textbf{Performance of backdoor task with different triggers. } The effectiveness of backdoor attacks varies significantly with different triggers across the datasets. For Fashion-MNIST, triggers $BA_6$ and $BA_7$ are highly effective, achieving backdoor accuracies of 96.90\% and 97.73\% respectively, while trigger $BA_8$ is notably less effective at 41.26\%. In MNIST, trigger $BA_1$ achieves near-perfect backdoor accuracy at 99.40\%, with most other triggers also showing high effectiveness except for $BA_8$, which has a lower accuracy of 49.30\%. This could be due to the shape of trigger 8, which has dimensions $24 \times 1$ and follows the vertical direction of the image, making it harder to detect in the image.
For CIFAR-10, trigger $BA_7$ is the most effective with a backdoor accuracy of 94.09\%, while $BA_8$ is the least effective at 27.19\%. Interestingly, all triggers show negligible impact in Tiny-ImageNet, with backdoor accuracy close to zero. This suggests that the dataset's complexity, potentially due to a smaller model size or a large number of classes, makes it challenging to learn both the main task and a backdoor task simultaneously. The presence of multiple backdoor tasks further complicates the attackers' efforts to embed the backdoor effectively.

\subsubsection{Semi-multiple-shot attack}

\begin{figure}[!hbt]
    \centering
    \includegraphics[width=1.0\linewidth]{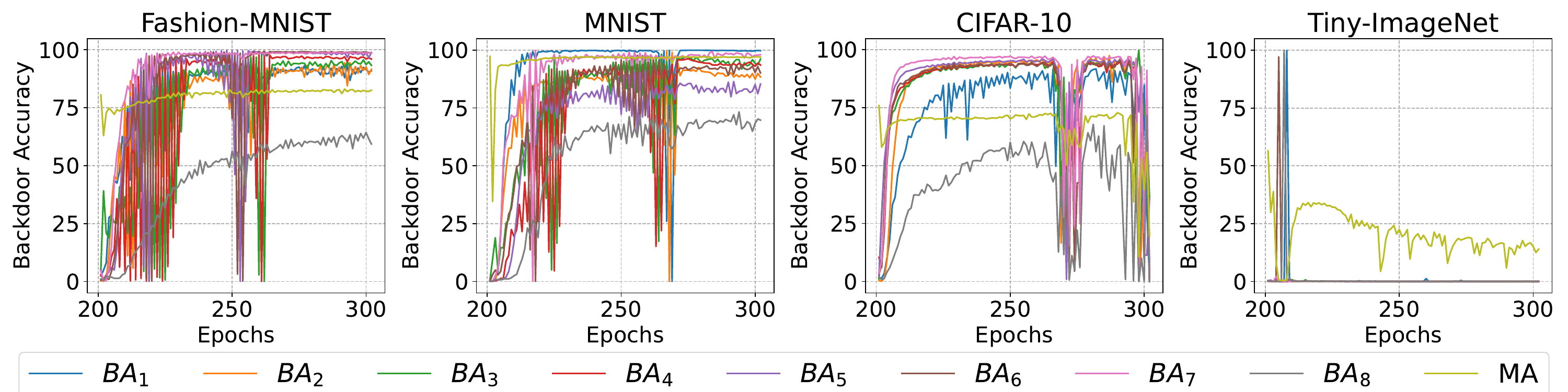}
    \caption{Backdoor accuracy in semi-multiple-shot NBA with $\gamma=\frac{100}{\#Atk}$}
    \label{fig:nba-semi-mul-1}
\end{figure}

\begin{table}[!hbt]
    \centering
    \caption{Performance of NBA in semi-multiple-shot scenario (8 adversaries) after 100 attack rounds}
    \label{tab:nba-semi-multiple-shot}
    \small
    \begin{tabular}{@{}lcccccccccc@{}}
    \toprule
 
\makecell{\textbf{Accuracy} $\rightarrow$,\\ \textbf{Dataset} $\downarrow$} & \textbf{$MA$} & \textbf{$BA_1$} & \textbf{$BA_2$} & \textbf{$BA_3$} & \textbf{$BA_4$} & \textbf{$BA_5$} & \textbf{$BA_6$} & \textbf{$BA_7$} & \textbf{$BA_8$} & \textbf{$BA_{Avg}$} \\ \midrule

    Fashion-MNIST & 82.53 & 90.93 & 91.54 & 93.29 & 96.01 & 98.79 & 98.77 & 98.68 & 59.33  & 90.92 \\
    MNIST & 97.04 & 99.76 & 88.38 & 96.10 & 93.55 & 85.32 & 90.15 & 97.99 & 69.74  & 90.12 \\
    CIFAR-10 & 19.71 & 0.01 & 0.00 & 35.74 & 0.00 & 35.94 & 0.00 & 0.00 & 0.00  & 8.96 \\
    Tiny-ImageNet & 13.95 & 0.00 & 0.00 & 0.00 & 0.00 & 0.00 & 0.00 & 0.00 & 0.00  & 0.00 \\
\bottomrule
\end{tabular}
\end{table}

In multiple-shot attacks, adversaries continuously inject backdoor triggers, resulting in a gradual increase in backdoor accuracy. As illustrated in Fig.~\ref{fig:nba-mul-1}, attackers need to participate in at least 200 rounds to achieve high backdoor accuracy for most triggers across three datasets, with the exception of Tiny-ImageNet. In real-world scenarios, however, adversaries may not always participate in every round. To reduce the number of rounds needed for effective attacks, we propose a semi-multiple-shot attack where adversaries participate for 100 rounds and then stop injecting backdoor triggers. The key difference between multiple-shot and semi-multiple-shot attacks is the adjustment of the scaling factor \(\gamma\) to \(\frac{100}{\#Atk}\), where \(\#Atk\) is the number of attackers. Results presented in Tab.~\ref{tab:nba-semi-multiple-shot} and Fig.~\ref{fig:nba-semi-mul-1} for a scenario with eight adversaries show that, although the backdoor accuracy fluctuates, it can reach high values for most triggers in three datasets at specific rounds.

These findings highlight a double-edged sword for backdoor attackers using the semi-multiple-shot attack. While it can be more efficient than the multiple-shot attack in terms of reducing participation rounds (reducing the risk of detection), it may not guarantee consistently high backdoor accuracy and the main accuracy might drop significantly. This inconsistency is crucial because once the backdoor injection is stopped, the model's backdoor accuracy (BA) drops significantly. However, the BAs then gradually recover, suggesting that the triggers are not entirely "forgotten" by the model. This presents a challenge for central server detection, as any client can inject a backdoor trigger in any round and then stop participating. The lingering effect of the triggers, even after the attacker ceases participation, makes it difficult for the server to distinguish between a temporary fluctuation and a true backdoor attack.

\subsection{The robustness in Non-Cooperative Backdoor Attacks}

\textbf{Norm Clipping~\cite{Sun2019CanYR}.} Participant updates undergo a clipping process to limit the impact of model adjustments, which involves multiplying them by $min(1, \frac{S}{\vert\vert L_{i}^{t + 1} - G_{t} \vert\vert}_2)$, where $S$ represents the clipping threshold. Tab.~\ref{tab:nba-multiple-shot} and Tab.~\ref{tab:nba-semi-multiple-shot-def-clip} illustrate the performance of the NBA algorithm in a multiple-shot scenario both before and after implementing the norm clipping defense with $S=5$. The results reveal that neither the primary accuracy nor the backdoor accuracy is significantly influenced by the norm clipping defense, suggesting that adversaries can still effectively introduce backdoor triggers into the global model.

\begin{table}[!hbt]
    \centering
    \caption{Performance of NBA in multiple-shot setting (8 adversaries) under norm clipping defense}
    \label{tab:nba-semi-multiple-shot-def-clip}
    \small
    \begin{tabular}{@{}lcccccccccc@{}}
    \toprule
 
\makecell{\textbf{Accuracy} $\rightarrow$,\\ \textbf{Dataset} $\downarrow$} & \textbf{$MA$} & \textbf{$BA_1$} & \textbf{$BA_2$} & \textbf{$BA_3$} & \textbf{$BA_4$} & \textbf{$BA_5$} & \textbf{$BA_6$} & \textbf{$BA_7$} & \textbf{$BA_8$} & \textbf{$BA_{Avg}$} \\ \midrule

Fashion-MNIST & 84.09 & 86.56 & 84.14 & 86.36 & 89.67 & 95.73 & 96.91 & 97.77 & 41.28 & 84.80 \\ 
MNIST & 97.95 & 99.39 & 86.62 & 89.44 & 87.62 & 79.36 & 89.42 & 96.95 & 48.94 & 84.72 \\
CIFAR-10 & 75.69 & 65.30 & 90.80 & 87.61 & 88.56 & 91.87 & 88.10 & 94.88 & 25.99 & 79.14 \\
Tiny-ImageNet & 42.53 & 0.03 & 0.02 & 0.08 & 0.06 & 0.05 & 0.04 & 0.02 & 0.13 & 0.05 \\

\bottomrule
\end{tabular}
\end{table}

\textbf{Differential Privacy (DP)~\cite{bagdasaryan2020backdoor}.} Gaussian noise $N(0, \sigma)$ is added to local updates to reduce the influence of backdoor attacks. As shown in Tab.~\ref{tab:nba-semi-multiple-shot-def-dp}, although the backdoor accuracy decreases, the DP defense significantly impacts the main accuracy, leading to a notable drop in performance across all datasets. This indicates a trade-off between maintaining privacy and preserving model efficacy.

\begin{table}[!hbt]
    \centering
    \caption{Performance of NBA in multiple-shot setting (8 adversaries) under DP defense}
    \label{tab:nba-semi-multiple-shot-def-dp}
    \small
    \begin{tabular}{@{}lcccccccccc@{}}
    \toprule
 
\makecell{\textbf{Accuracy} $\rightarrow$,\\ \textbf{Dataset} $\downarrow$} & \textbf{$MA$} & \textbf{$BA_1$} & \textbf{$BA_2$} & \textbf{$BA_3$} & \textbf{$BA_4$} & \textbf{$BA_5$} & \textbf{$BA_6$} & \textbf{$BA_7$} & \textbf{$BA_8$} & \textbf{$BA_{Avg}$} \\ \midrule

Fashion-MNIST & 82.67 & 85.00 & 78.69 & 83.23 & 91.92 & 89.19 & 94.67 & 97.91 & 33.36 & 81.75 \\
MNIST & 97.02 & 95.64 & 89.28 & 70.38 & 88.09 & 70.96 & 84.48 & 95.01 & 44.64 & 79.81 \\
CIFAR-10 & 41.66 & 59.43 & 72.26 & 52.92 & 65.04 & 78.11 & 43.52 & 65.38 & 12.98 & 56.21 \\
Tiny-ImageNet & 9.38 & 0.02 & 0.00 & 0.35 & 0.00 & 0.08 & 0.09 & 0.04 & 0.53 & 0.14 \\

\bottomrule
\end{tabular}
\end{table}

\textbf{Effectiveness of defense mechanisms.} Mainstream defenses, such as Norm Clipping and Differential Privacy, were not designed to address scenarios where multiple independent attackers inject unique triggers and target classes. Additionally, existing research hasn't thoroughly investigated the effectiveness of these defenses in the presence of multiple backdoor attackers (NBA). This highlights a critical gap in the current understanding of defense mechanisms for FL security.

\subsection{Limitations}
\label{subsec:limitations-discussion}
Our work explores various facets of the NBA scenario in FL systems, focusing on trigger design, dataset characteristics, and model updates. The effectiveness of backdoor triggers varies with dataset characteristics, and the scaling factor $\gamma$ significantly impacts backdoor accuracy and main task performance. Multiple-shot attacks require numerous rounds to achieve high backdoor accuracy, which is impractical in real-world scenarios, while semi-multiple-shot attacks can reduce the number of rounds but need precise tuning of $\gamma$. Our controlled evaluation might not fully capture real-world FL complexities, and our study focuses on limited trigger designs and defense mechanisms like norm clipping and differential privacy. Future research should explore more sophisticated attack strategies and novel defenses tailored to the non-i.i.d. nature of FL systems.

\section{Conclusion}
This paper explores the emerging threat of the NBA scenario in FL systems, where multiple independent clients can compromise the global model by injecting unique backdoor triggers and target classes. Our findings highlight the potential of watermarking-based backdoor triggers, which can be useful in cross-silo FL scenarios to protect the copyright of participants. This study lays the groundwork for future research focused on developing robust strategies to counteract backdoor attacks. Furthermore, investigating incentive structures that discourage malicious behavior and encourage cooperative participation is essential. By advancing in these areas, we can significantly improve the security, privacy, and integrity of FL platforms, thereby contributing to the creation of secure and trustworthy FL systems.

\bibliographystyle{unsrtnat} 
\bibliography{ref} 
\newpage
\appendix
This appendix provides an extended exploration of our research, providing additional details on methods and results. Appx.~\ref{appsub:exp_setup} details the training procedures and experimental settings used in our NBA experiments. Appx.~\ref{appsub:main-one-atk} presents additional results on NBA performance with a single attacker. We then explore the impact of multiple attackers on NBA performance in Appx.~\ref{appsub:multi-atk}, showcasing results with varying attacker counts. Appx.~\ref{appsub:multi-atk-def} evaluates the effectiveness of defense mechanisms against NBA attacks, offering additional results. Finally, Appx.~\ref{appsub:discussion} and Appx.~\ref{appsub:societal} discuss the limitations of our work, explore potential social implications, and suggest future research directions. 

\section{Training details and experimental settings in NBA}
\label{appsub:exp_setup}

\subsection{Experiment setup}

\textbf{Datasets and hyperparameters.} We use four datasets in our experiments: Fashion-MNIST~\cite{xiao2017fashion}, MNIST~\cite{lecun1998gradient}, CIFAR-10~\cite{krizhevsky2009learning}, and Tiny-ImageNet~\cite{le2015tiny}. These datasets are preprocessed and divided among different participants in the FL system with heterogeneous data distributions, set at $\alpha = 0.5$. To simulate the NBA scenario, we use a total of $N = 100$ clients, with $K = 10$ clients selected for aggregation in each round. During each round, each client trains for $E$ local epochs with a local learning rate $lr$ and a batch size of 128. The global learning rate $\eta$ is set to 0.1. The details of the FL training setup are shown in Table~\ref{tab:exp-setup}.

\begin{table}[!hbt]
\centering
\caption{NBA training details}  
\label{tab:exp-setup}
\begin{tabular}{lccccc}
\toprule
\textbf{Dataset} & \textbf{Model used} & \#Client ($K / N$) & Benign $lr / E$ & Poison $lr / E$ & $\eta$ \\ \midrule
Fashion-MNIST & 2 conv and 2 fc & 10 / 100 & 0.1 / 2 & 0.05 / 6 & 0.1 \\
MNIST & 2 conv and 2 fc & 10 / 100 & 0.1 / 2 & 0.05 / 6 & 0.1 \\
CIFAR-10 & Resnet-18~\cite{he2016deep} & 10 / 100 & 0.1 / 2 & 0.05 / 6 & 0.1 \\
Tiny-ImageNet & Resnet-18~\cite{he2016deep} & 10 / 100 & 0.1 / 2 & 0.05 / 6 & 0.1 \\
\bottomrule
\end{tabular}
\end{table}

\textbf{Computational Resources.} All experiments are conducted on a server with an Intel(R) Xeon(R) Gold 6242 CPU @ 2.80GHz, 256GB RAM, and an NVIDIA GeForce RTX 3090 GPU with 24GB memory.
The code is implemented in PyTorch~\cite{paszke2019pytorch} and the experiments are conducted following the backdoor attack setup in prior work~\cite{bagdasaryan2020backdoor, xie2020dba}.
Moreover, we utilized libraries such as NumPy~\cite{harris2020array}, and Matplotlib~\cite{hunter2007matplotlib} for data processing and visualization.

\subsection{Model Replacement Attack}
In Model Replacement Attack~\cite{bagdasaryan2020backdoor}, the adversary aims to completely overwrite the global model $G^{t+1}$ with their malicious model, denoted by $X$ using the following equation:
\begin{equation}
X = G^{t} + \frac{\eta}{n}\sum_{i=1}^m (W_i^{t+1} - G^t).  \tag{4}
\end{equation}
Due to the non-independent and identically distributed (non-IID) nature of the training data, local models ($W_i^{t+1}$ ) may deviate significantly from the current global model ($G^t$). However, as the global model converges, these deviations tend to cancel each other out, i.e., $\sum_{i=1}^{m-1} (W_i^{t+1} -
G^t) \approx 0$, meaning the sum of these deviations approaches zero.

Leveraging this cancellation effect, the adversary can solve for the malicious model they need to submit ($\widetilde{W}_m^{t+1}$ ) by:
\begin{equation}
    \widetilde{W}_m^{t+1} =
    \frac{n}{\eta}X - (\frac{n}{\eta}-1)G^{t} - \sum_{i=1}^{m-1}(W_i^{t+1} - G^t)
    \approx
    \frac{n}{\eta}(X - G^t) + G^{t}.  \tag{5}
    \label{eq:scaling}
\end{equation}

\section{Performance of main task with one adversary}
\label{appsub:main-one-atk}

\subsection{Single-shot attack}
\begin{figure}[!hbt]
    \centering
    \includegraphics[width=1.0\linewidth]{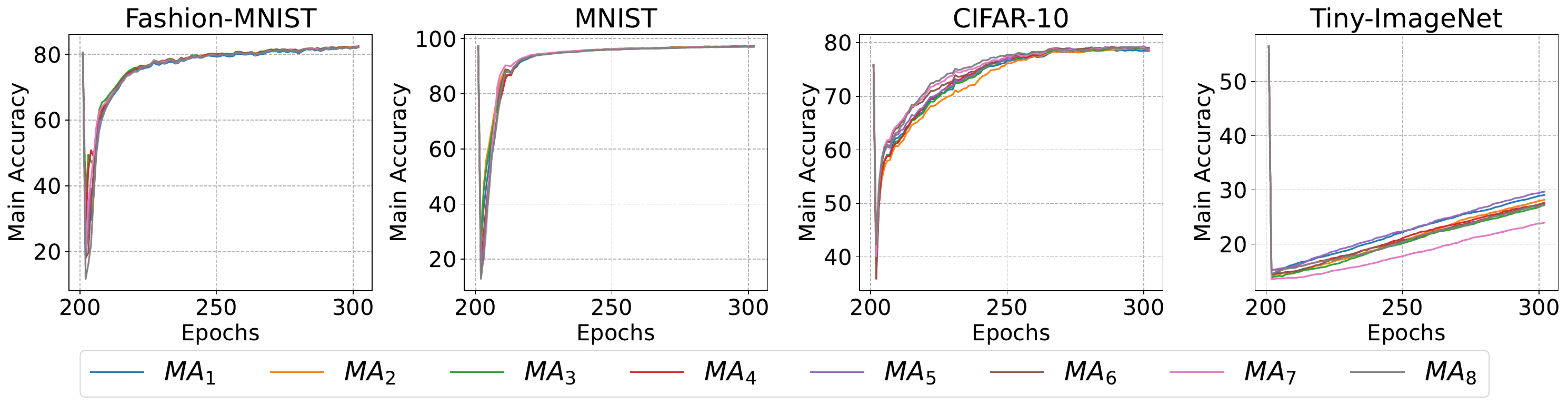}
    \caption{Main accuracy of single-shot attack with one adversary}
    \label{fig:main-one-single}
\end{figure}

The introduction of backdoor attacks in single-shot scenarios caused a substantial drop in main accuracy across all datasets during the attack round, with the most pronounced impacts observed in Tiny-ImageNet and MNIST, as illustrated in Tab.~\ref{tab:main-acc-singshot} and Fig.~\ref{fig:main-one-single}, indicating their higher susceptibility. Post-attack, the main accuracy largely recovered in Fashion-MNIST, MNIST, and CIFAR-10, suggesting that these models can regain performance with continued training. However, Tiny-ImageNet's main accuracy remained significantly lower than the initial stage even after 100 training rounds, indicating a more lasting impact or greater difficulty in overcoming the attack. The consistent patterns across different triggers, with similar recovery trends in main accuracy, highlight the varying resilience of datasets to backdoor attacks in FL and underscore the need for tailored defense mechanisms to mitigate long-term impacts effectively.

\begin{table}[!hbt]
    \centering
    \caption{Performance of the main task with one adversary in single-shot after 100 rounds ($\gamma=100$)}
    \label{tab:main-acc-singshot}
    \small
    \begin{tabular}{@{}lccccccccc@{}}
    \toprule
 
\makecell{\textbf{Accuracy} $\rightarrow$,\\ \textbf{Dataset} $\downarrow$} & \textbf{$MA_1$} & \textbf{$MA_2$} & \textbf{$MA_3$} & \textbf{$MA_4$} & \textbf{$MA_5$} & \textbf{$MA_6$} & \textbf{$MA_7$} & \textbf{$MA_8$} & \textbf{$MA_{Avg}$} \\ \midrule

    Fashion-MNIST & 82.15 & 82.51 & 82.30 & 82.49 & 82.38 & 82.29 & 82.28 & 82.19 & 82.32 \\
    MNIST & 97.12 & 97.23 & 97.25 & 97.15 & 97.12 & 97.13 & 97.07 & 97.00 & 97.13 \\
     CIFAR-10 & 78.50 & 78.67 & 79.03 & 79.02 & 79.10 & 78.95 & 78.91 & 78.84 & 78.88 \\
    Tiny-ImageNet & 29.04 & 28.16 & 27.22 & 27.64 & 29.66 & 27.49 & 23.95 & 27.20 & 27.54 \\
    \midrule

\end{tabular}
\end{table}
\label{appsub:single-shot-one-atk}

\subsection{Multiple-shot attack}
\label{appsub:multiple-shot-one-atk}

\begin{table}[!hbt]
    \centering
    \caption{Performance of the main task with one adversary in multiple-shot setting}
    \label{tab:main-acc-multipleshot}
    \small
    \begin{tabular}{@{}lccccccccc@{}}
    \toprule
 
\makecell{\textbf{Accuracy} $\rightarrow$,\\ \textbf{Dataset} $\downarrow$} & \textbf{$MA_1$} & \textbf{$MA_2$} & \textbf{$MA_3$} & \textbf{$MA_4$} & \textbf{$MA_5$} & \textbf{$MA_6$} & \textbf{$MA_7$} & \textbf{$MA_8$} & \textbf{$MA_{Avg}$} \\ \midrule

    Fashion-MNIST & 87.50 & 87.34 & 87.46 & 87.56 & 87.45 & 87.45 & 87.54 & 87.58 & 87.48 \\
    MNIST & 98.82 & 98.87 & 98.86 & 98.88 & 98.86 & 98.86 & 98.92 & 98.83 & 98.86 \\
    CIFAR-10 & 75.97 & 77.04 & 77.34 & 77.79 & 77.79 & 77.74 & 77.86 & 76.40 & 77.24 \\
    Tiny-ImageNet & 52.03 & 52.85 & 53.09 & 53.20 & 52.99 & 53.40 & 52.94 & 52.40 & 52.86 \\
    \midrule

\end{tabular}
\end{table}

\begin{figure}[!hbt]
    \centering
    \includegraphics[width=1.0\linewidth]{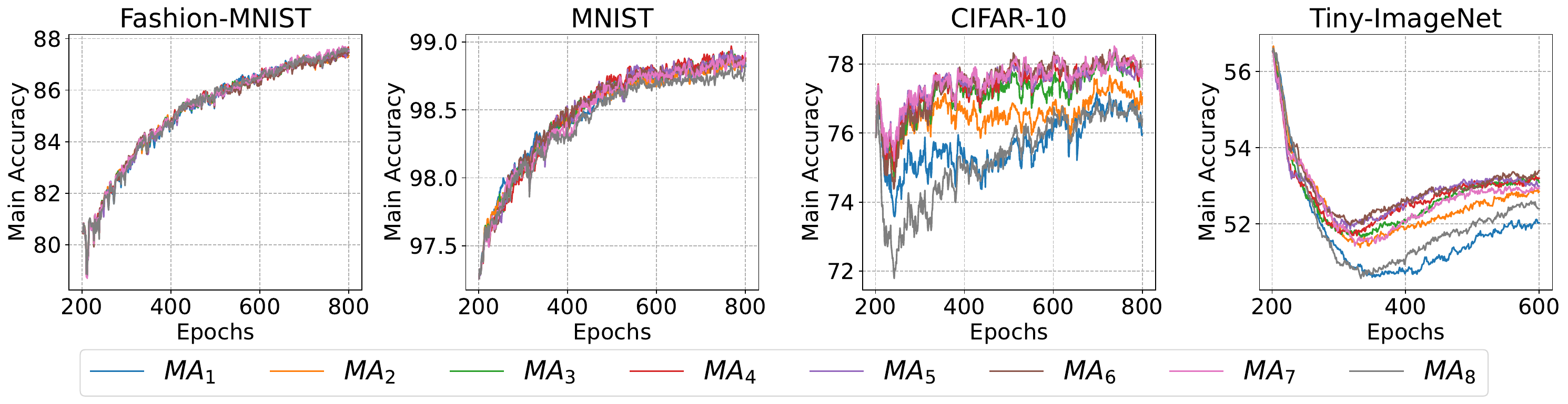}
    \caption{Main task accuracy of multiple-shot setting with one adversary}
    \label{fig:main-one-multiple}
\end{figure}

In the multiple-shot attack scenario, the main accuracy, as shown in Tab.~\ref{tab:main-acc-multipleshot} and Fig.~\ref{fig:main-one-multiple}, generally improved or remained stable after the attack period compared to the initial accuracy. Both Fashion-MNIST and MNIST showed an increase in main accuracy, suggesting that the models could recover and even benefit from continued training despite the attacks. CIFAR-10 experienced a slight improvement, indicating minimal negative impact and marginal gains. However, Tiny-ImageNet saw a slight decrease in main accuracy, highlighting its continued susceptibility to backdoor attacks. Overall, most datasets could recover or improve following the attack, except for Tiny-ImageNet, which underscores the need for stronger defense mechanisms in FL to protect more complex datasets.

\section{Non-Cooperative Backdoor Attacks with different numbers of adversaries}
\label{appsub:multi-atk}
\begin{table}[!hbt]
    \centering
    \caption{Performance of NBA in multiple-shot ($\gamma = 1$) with different numbers of adversaries}
    \label{tab:ba-multiple-shot}
    \small
    \begin{tabular}{@{}lccccccccccc@{}}
    \toprule
 
\makecell{\textbf{Accuracy} $\rightarrow$,\\ \textbf{Dataset} $\downarrow$} & \textbf{\#Atk} & \textbf{$MA$} & \textbf{$BA_1$} & \textbf{$BA_2$} & \textbf{$BA_3$} & \textbf{$BA_4$} & \textbf{$BA_5$} & \textbf{$BA_6$} & \textbf{$BA_7$} & \textbf{$BA_8$} & \textbf{$BA_{Avg}$} \\ \midrule

\multirow{8}{*}{\makecell{Fashion-\\MNIST}}
     & 1 & 86.58 & 97.38  & -  & -  & -  & -  & -  & -  & -  & 97.38 \\
     & 2 & 86.13 & 86.31 & 82.44  & -  & -  & -  & -  & -  & -  & 84.38 \\
     & 3 & 85.78 & 88.83 & 83.23 & 79.68  & -  & -  & -  & -  & -  & 83.91 \\
     & 4 & 84.92 & 88.19 & 84.74 & 78.34 & 74.68  & -  & -  & -  & -  & 81.49 \\
     & 5 & 85.00 & 87.86 & 84.07 & 85.04 & 84.32 & 97.44  & -  & -  & -  & 87.75 \\
     & 6 & 84.61 & 87.53 & 85.03 & 87.93 & 88.52 & 93.72 & 97.50  & -  & -  & 90.04 \\
     & 7 & 84.30 & 86.63 & 82.82 & 86.21 & 89.48 & 95.18 & 95.62 & 97.22  & -  & 90.45 \\
     & 8 & 84.07 & 86.53 & 84.11 & 86.34 & 89.68 & 95.71 & 96.90 & 97.73 & 41.26  & 84.78 \\
\midrule

\multirow{8}{*}{\makecell{MNIST}}
     & 1 & 98.72 & 96.82  & -  & -  & -  & -  & -  & -  & -  & 96.82 \\
     & 2 & 98.72 & 96.40 & 83.60  & -  & -  & -  & -  & -  & -  & 90.00 \\
     & 3 & 98.65 & 99.49 & 87.07 & 86.63  & -  & -  & -  & -  & -  & 91.06 \\
     & 4 & 98.49 & 99.30 & 88.82 & 78.19 & 52.97  & -  & -  & -  & -  & 79.82 \\
     & 5 & 98.42 & 99.28 & 87.23 & 87.53 & 76.99 & 88.77  & -  & -  & -  & 87.96 \\
     & 6 & 98.31 & 99.18 & 86.43 & 91.58 & 80.41 & 83.68 & 88.93  & -  & -  & 88.37 \\
     & 7 & 98.07 & 99.07 & 86.46 & 89.75 & 84.65 & 75.97 & 89.20 & 91.40  & -  & 88.07 \\
     & 8 & 97.96 & 99.40 & 86.61 & 89.47 & 87.52 & 79.24 & 89.43 & 97.01 & 49.30  & 84.75 \\

\midrule

\multirow{8}{*}{\makecell{CIFAR-10}}
     & 1 & 76.04 & 61.33  & -  & -  & -  & -  & -  & -  & -  & 61.33 \\
     & 2 & 75.04 & 45.36 & 84.32  & -  & -  & -  & -  & -  & -  & 64.84 \\
     & 3 & 75.34 & 31.97 & 81.30 & 78.18  & -  & -  & -  & -  & -  & 63.81 \\
     & 4 & 75.77 & 41.00 & 85.97 & 78.82 & 80.20  & -  & -  & -  & -  & 71.50 \\
     & 5 & 76.29 & 48.38 & 86.89 & 81.66 & 82.53 & 88.99  & -  & -  & -  & 77.69 \\
     & 6 & 75.63 & 48.69 & 87.52 & 81.61 & 84.81 & 89.81 & 85.58  & -  & -  & 79.67 \\
     & 7 & 76.62 & 46.60 & 87.82 & 79.88 & 83.86 & 89.18 & 80.01 & 92.76  & -  & 80.01 \\
     & 8 & 76.21 & 55.87 & 88.93 & 84.69 & 86.78 & 89.99 & 86.50 & 94.09 & 27.19  & 76.75 \\

\midrule

\multirow{8}{*}{\makecell{Tiny-\\ImageNet}}
     & 1 & 51.54 & 93.55  & -  & -  & -  & -  & -  & -  & -  & 93.55 \\
     & 2 & 47.35 & 89.29 & 86.29  & -  & -  & -  & -  & -  & -  & 87.79 \\
     & 3 & 45.19 & 89.91 & 4.48 & 1.94  & -  & -  & -  & -  & -  & 32.11 \\
     & 4 & 43.75 & 77.42 & 1.66 & 0.20 & 2.39  & -  & -  & -  & -  & 20.42 \\
     & 5 & 42.35 & 54.23 & 6.13 & 1.64 & 8.68 & 2.52  & -  & -  & -  & 14.64 \\
     & 6 & 41.16 & 24.55 & 2.66 & 0.98 & 2.35 & 0.32 & 1.71  & -  & -  & 5.43 \\
     & 7 & 40.66 & 0.44 & 0.04 & 0.10 & 0.10 & 0.01 & 0.03 & 0.01  & -  & 0.10 \\
     & 8 & 38.32 & 0.01 & 0.00 & 0.05 & 0.03 & 0.00 & 0.00 & 0.00 & 0.04  & 0.02 \\

\midrule

\multicolumn{11}{l}{\textit{Note:} \textbf{\#Atk} denotes the number of adversaries.}\\

\end{tabular}
\end{table}

In the multiple-shot attack scenario with varying numbers of adversaries, the impact on main and backdoor accuracies across different datasets reveals distinct trends. While Fashion-MNIST and MNIST exhibit relatively stable main accuracies as the number of adversaries increases, indicating resilience to the attacks, the effectiveness of the backdoor attacks diminishes, as reflected by decreasing average backdoor accuracies. In CIFAR-10, the main accuracy remains consistent, but the backdoor accuracy shows an upward trend with more adversaries, suggesting increased susceptibility to backdoor attacks. In contrast, Tiny-ImageNet experiences significant drops in both main and backdoor accuracies with an increasing number of adversaries, indicating heightened vulnerability. These findings underscore the importance of considering the number of adversaries when designing defense mechanisms in federated learning systems, as different datasets may exhibit varying levels of resilience to multiple adversaries (see Tab.~\ref{tab:ba-multiple-shot}).

\section{NBA with eight adversaries in multiple-shot setting under different defenses}
\label{appsub:multi-atk-def}
\begin{figure}[!hbt]
    \centering
    \includegraphics[width=1.0\linewidth]{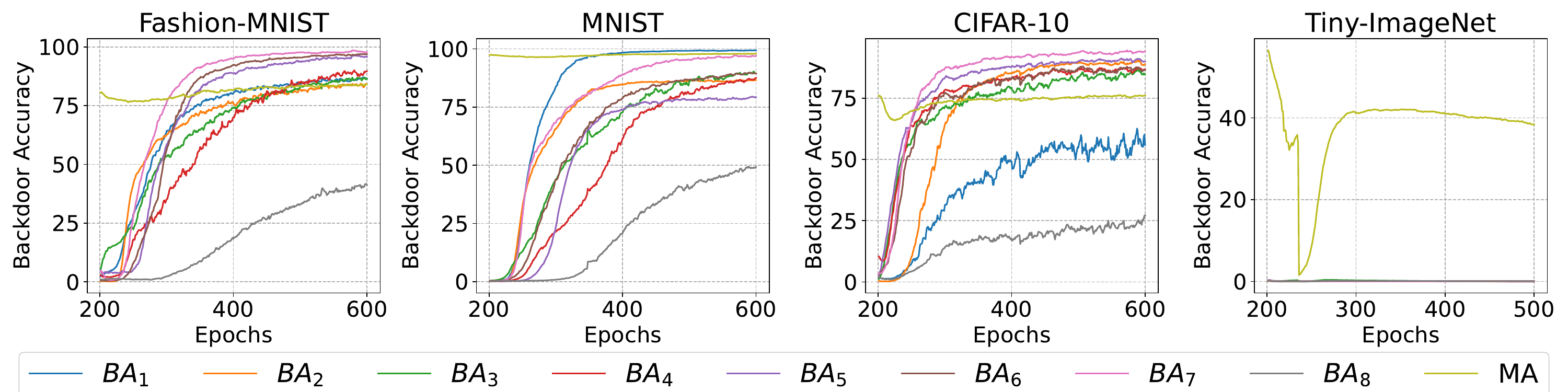}
    \caption{NBA with eight adversaries in multiple-shot setting without defense}
    \label{fig:nba-def-wdef}
\end{figure}

\begin{figure}[!hbt]
    \centering
    \includegraphics[width=1.0\linewidth]{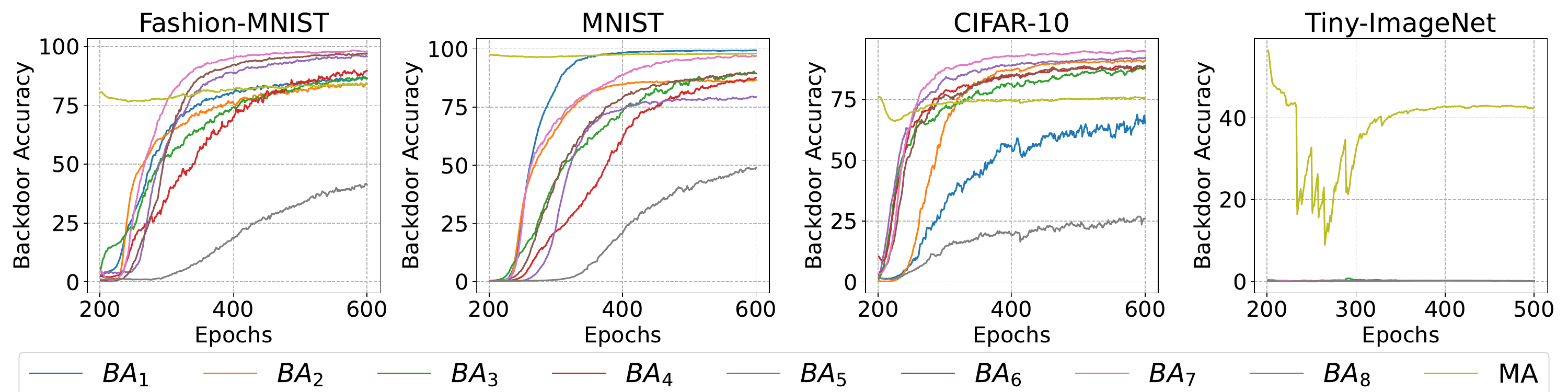}
    \caption{NBA with eight adversaries in multiple-shot setting under Norm Clipping defense}
    \label{fig:nba-def-wclip}
\end{figure}

\begin{figure}[!hbt]
    \centering
    \includegraphics[width=1.0\linewidth]{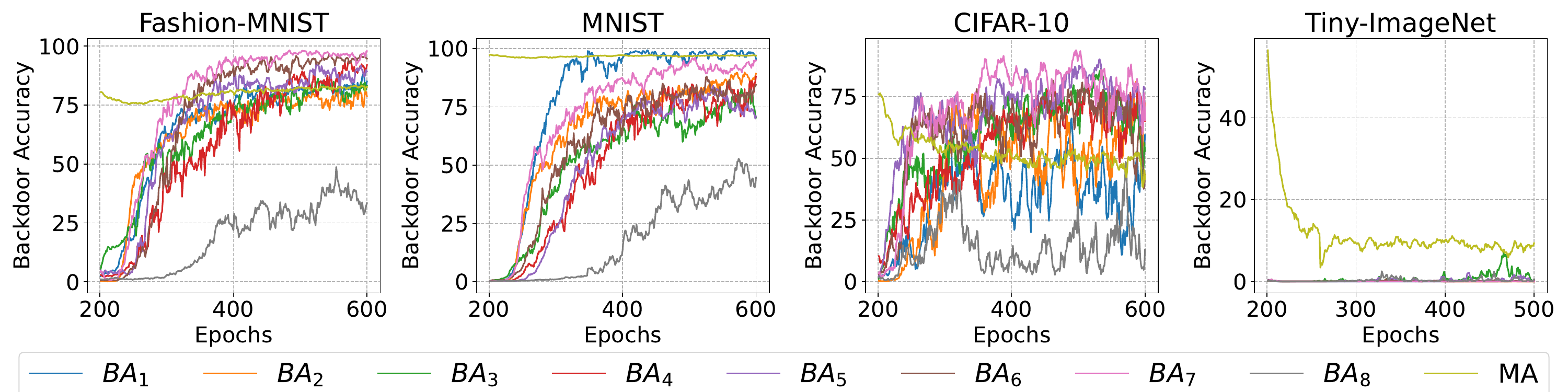}
    \caption{NBA with eight adversaries in multiple-shot setting under Differential Privacy defense}
    \label{fig:nba-def-wdp}
\end{figure}

In the scenario of multiple-shot attacks involving eight adversaries, the effects on main and backdoor accuracies across various datasets display discernible patterns. Without defense mechanisms (see Figure \ref{fig:nba-def-wdef}), the model's accuracies vary across datasets, with Fashion-MNIST and MNIST exhibiting relatively higher main accuracies compared to CIFAR-10 and Tiny-ImageNet. However, all datasets show significant backdoor accuracies, indicating vulnerability to backdoor attacks. When employing the Norm Clipping defense method (see Figure \ref{fig:nba-def-wclip}), there are slight improvements in main accuracies across all datasets, with Fashion-MNIST and MNIST maintaining similar backdoor accuracies while CIFAR-10 and Tiny-ImageNet experience slight decreases in backdoor accuracies. The Norm Clipping defense appears to be more effective in mitigating backdoor attacks in CIFAR-10 and Tiny-ImageNet compared to Fashion-MNIST and MNIST. On the other hand, employing the Differential Privacy defense method (see Figure \ref{fig:nba-def-wdp}) results in noticeable drops in both main and backdoor accuracies across all datasets. This indicates that while Differential Privacy may offer some protection against backdoor attacks, it also significantly impacts the model's overall performance, particularly in CIFAR-10 and Tiny-ImageNet, where main accuracies decrease considerably. Therefore, the choice of defense mechanism should consider the trade-off between backdoor protection and maintaining overall model performance.

\section{Discussion}
\label{appsub:discussion} 
Our work delves into multiple aspects of the NBA scenario within FL systems, particularly emphasizing trigger design, dataset characteristics, and model updates. The effectiveness of backdoor triggers can vary widely depending on the dataset; some datasets may render the trigger ineffective, while others may facilitate highly effective backdoor attacks. Additionally, the scaling factor $\gamma$ plays a critical role in the success rate of these attacks. Increasing the scale of model updates can significantly enhance backdoor accuracy, yet this comes at the cost of potentially degrading the main task accuracy, leading to suboptimal overall performance.

In the context of multiple-shot attacks, adversaries must engage in a substantial number of rounds to achieve high backdoor accuracy, which may be impractical in real-world scenarios. However, in cross-silo FL scenarios, each participant can introduce unique backdoor triggers as a form of watermarking to protect their intellectual property. Semi-multiple-shot attacks offer a more efficient alternative by reducing the number of rounds required to attain high backdoor accuracy, though they require careful selection of the scaling factor $\gamma$ to ensure effectiveness and may not consistently maintain high accuracy throughout the attack.

\section{Societal impacts}
\label{appsub:societal} 
Our research highlights the potential of NBA to compromise the integrity of FL systems. We believe this work serves as a crucial stepping stone towards a more secure future for FL. Through our work, we highlight the efficacy of watermarking-based triggers, presenting avenues for enhancing secure communication and detecting tampering within FL frameworks. Additionally, we explore the design of incentive structures, aiming to cultivate cooperative engagement while mitigating malicious activities. These endeavors are pivotal in establishing trust and fostering a secure ecosystem within FL platforms.
By advancing research in these areas, we contribute substantially to the establishment of robust and trustworthy FL systems. Such efforts are crucial for safeguarding data privacy and integrity, thereby unleashing the full potential of FL to benefit society at large.

\end{document}